# Observation of plasmonic exceptional points

J.-H. Park[1,4]*, A. Ndao[1,4]*, W. Cai[4], L.-Y. Hsu[1,4], A. Kodigala[4], T. Lepetit[4], Y.-H. Lo[4], and B. Kanté[1,2,3,4]

[1]*Department of Electrical Engineering and Computer Sciences, University of California, Berkeley, California 94720, USA*
[2]*Department of Mechanical Engineering, University of California, Berkeley, California 94720, USA*
[3]*Materials Sciences Division, Lawrence Berkeley National Laboratory, 1 Cyclotron Road, Berkeley, California 94720, USA*
[4]*Department of Electrical and Computer Engineering, University of California San Diego, La Jolla, CA 92093-0407, USA*
bkante@berkeley.edu
*These authors contributed equally to this work

**Abstract**- Non-Hermitian singularities known as exceptional-points (EPs) have been shown to exhibit increased sensitivities but the observation of EPs has so far been limited to wavelength scaled systems subject to diffraction limit. We propose a novel approach to EPs and report their first observation in plasmonics at room temperature. The plasmonic EPs are based on the hybridization of detuned resonances in multilayered plasmonic crystals to reach a critical complex coupling rate between nanoantennas arrays, and, resulting in the simultaneous coalescence of the resonances and loss rates. Because plasmons shrink the wavelength of light to make it compatible with biological relevant substances, enhanced sensing of anti-Immunoglobulin G, the most common antibody found in blood circulation, is observed. Our work opens the way to novel class of nanoscale devices, sensors, and imagers based on topological polaritonic effects.

Sensing is fundamental to our observation of the universe via physical quantities such as mass, time, or distances. Optical molecular nanosensing, i.e., the ability to detect extremely small quantities with light, enables the detection of threats at early stage and will revolutionize security and medicine. Sensing technologies in classical and quantum regimes are usually based on non-destructive probing utilizing enhanced wave-matter interaction at resonances [1]. The interaction of waves with a sensor thus requires the latter to be an open system, i.e., a non-Hermitian system described by both radiative and absorptive processes. Recently, non-Hermitian singularities known as exceptional points (EPs) have been observed in systems including electromagnetism, atom-cavity, and acoustics [2-10]. EPs are singularities where at least two eigenmodes of an open system coalesce to become degenerate both in their resonance frequencies and loss rates. At such singularities, the topology of the system is drastically modified, and it appears skewed with reduced dimensionality but enhanced sensitivity [11]. To date, the observation of EPs has been restricted to wavelength scaled systems based on dielectric waveguides and resonators subject to diffraction limit [12-22]. While parity-time symmetry prescribes a systematic recipe to implement EPs in those systems, its implementation in subwavelength scale systems such as plasmonics, constitutes a formidable challenge requiring the controlled spatial distribution of loss and gain at extremely small scales beyond current nanofabrication capabilities [23-25]. The observation of such non-Hermitian singularities in plasmonics has thus remained elusive and has hampered the investigation of the physics of EPs at electronic or molecular scales [26]. Here, we report the first observation of EPs in plasmonics. The plasmonic EP is based on the hybridization of detuned resonators in a multilayered plasmonic crystal to reach a critical complex coupling rate resulting in the simultaneous coalescence of resonances and loss rates. We also demonstrate that the plasmonic EP enables enhanced sensing of anti-Immunoglobulin G, the most common antibody found in blood circulation. Our work opens the way to novel class of compact nanoscale sensors and imagers based on topological polaritonic effects.

Exceptional points exist in open multimode systems and the challenge of their observation resides in the identification of real physical parameters, i.e., simple geometrical parameters, enabling their implementation. Our system consists of a bilayer plasmonic crystal made of two optically dissimilar plasmonic resonators array with detuned resonances. The detuning can be implemented either using identical resonators in distinct optical environments (Fig. 1) or using structures of distinct size in a uniform optical environment (Fig. 2). In the schematic of Fig. 1a, one of the resonators is on glass and embedded in a 100 nm thick polymer spacer (SU-8) while the second resonator is on top of the polymer and is exposed to air, making its surface available for analytes binding. The gold metallic bars have a length l = 250 nm, a width w = 50 nm, and a thickness of t = 40 nm. The structures are fabricated using two steps electron beam lithography (EBL) and metal lift-off on a glass substrate ($n_{sub}$ = 1.5). The metal array patterns were defined in the bilayer e-beam resists with 170 nm thick methyl methacrylate and 50 nm thick polymethyl methacrylate with high-resolution EBL (EBPG5200, Vistec) followed by a 3 nm chromium and a 37 nm gold deposition using electron beam evaporation and a lift-off process. The chromium layer is used as an adhesion layer between glass and gold. Then, a 100 nm thick SU-8 2000.1 (MicroChem) is spun over on the first metal array as a planarized dielectric layer ($n_{SU-8}$ = 1.57). The roughness of the surface after planarization is under 3 nm (see supplementary information). Finally, the second layer was fabricated using the same method but including a precise alignment process. Figure 1b (left) presents a top-view scanning electron micrograph (SEM) of the fabricated multilayer structure, with a lateral shift between bars dx = 100 nm. Figure 1b (right) shows zoom-in top (XY-plane) and side (XZ-plane) views of the plasmonic crystal, clearly showing the top and bottom metallic bars and the quality of the fabrication and alignment processes. The side view image is obtained using a dual-beam focused ion beam (FIB)-SEM that simultaneously enables the local sectioning (with the FIB) and imaging (with SEM) of the samples. Px and Py are in-plane periodicities and dx is the lateral shift between the center of the bars along the direction of their electric dipolar mode (X-direction). The period along the X-direction is fixed to Px = 400 nm while Py and dx are the two parameters used to tune the coupling between resonators array to reach an exceptional point.

Figure 2 presents the real and imaginary parts of the eigenmodes of hybridized plasmonic resonators array of optically identical (Fig. 2a-b) and optically dissimilar structures as a function of the lateral shift between the center of the dipoles (dx) and the periodicity perpendicular to the electric dipole moment (Py). The optically dissimilar structures are implemented using either resonators of dissimilar size (Fig. 2c-d) embedded in a dielectric slab ($n_{slab}$ = 1.5 and $h_{z,slab}$ = 240 nm) or using resonators of identical size in distinct optical environments (Fig. 2e-f). The hybridization of optically identical resonators leads to symmetric and antisymmetric modes, and, resonances cross along a diabolic line as a function of dx and Py (Fig. 2a). Because of the opposite symmetry of the hybridized modes, their loss rates (Fig. 2b) are very different and are always avoided. The symmetric configuration can thus not lead to exceptional points (EPs) [27-29] and leads to usual Fano resonances where two modes of distinct loss rates overlap [30]. The hybridization of optically dissimilar resonators, however, leads to two hybrid modes with crossing and avoided crossing of both the resonances and loss rates (Fig. 2c-d, 2e-f) unambiguously demonstrating the existence of a plasmonic exceptional point where resonances and loss rates become simultaneously degenerate. It is worth noting that the loss rate for plasmonic EPs includes losses by radiation and absorption. The EP singularity (black dot) occurs at ~241 THz for dx ~ 120 nm and Py ~ 433 nm. The interplay between near-field Coulomb interactions and radiative coupling via interferences enables the coalescence of the hybrid modes. Materials parameters and numerical simulation details are presented in the supplementary information.

To further investigate the topology of the plasmonic EP, we analyze the dispersion of plasmonic modes around the singularity (~243 THz for dx of ~134 nm and for Py between 400 nm and 430 nm). Figure 2g presents numerical simulations of structures supporting a plasmonic EP and DP with a cladding layer described by a varying refractive index. The figure on the left presents the structure (nanosensor) with the cladding layer constituting the perturbation δ. The nanosensor operating around an EP (red line) exhibits resonance splitting (Δω) proportional to the square-root of the perturbation $\boldsymbol{\Delta\omega_{EP} \sim \sqrt{\delta}}$ whereas the DP nanosensor (blue line) exhibits resonance splitting that depends linearly on the perturbation $\boldsymbol{\Delta\omega_{DP} \sim \delta}$. The power laws are confirmed by log plots in Fig. 2h with slopes of 0.5 and 1.0 for the EP (Px = 400 nm, Py = 415 nm, dx= ∼134 nm) and DP (Px = 400 nm, Py = 350 nm dx= ∼161 nm) nanosensors respectively. The square root dependence on the perturbation constitutes an additional evidence of the successful implementation of a plasmonic EP.

To experimentally demonstrate plasmonic exceptional points, we characterize the structures of Fig. 1 as a function of the lateral shift dx for two different values of the period in the Y direction, Py = 400 nm and Py = 430 nm. Figure 3 presents experimental (circles) and simulated (dashed curves) resonance frequencies (ω) and loss rates (γ), extracted from complex scattering parameters (see supplementary information). A very good agreement is obtained. In Fig. 3a-b, we observe a crossing of resonance frequencies ($\omega_A$ and $\omega_B$) and an avoided crossing of loss rates ($\gamma_A$ and $\gamma_B$) for Py = 400 nm. In Fig. 3c-d, an avoided crossing of resonance frequencies ($\omega_A$ and $\omega_B$) and a crossing of loss rates ($\gamma_A$ and $\gamma_B$) are observed for Py = 430 nm. An EP singularity thus unambiguously occurs around ∼243 THz for dx of ∼134 nm and for Py between 400 nm and 430 nm. The experimental results are obtained by measuring both the amplitude and the phase of the transmitted light. The transmittance is measured using a Fourier-transform infrared spectrometer (Bruker Vertex 70) combined with an infrared microscope (×15 Cassegrain objective, numerical aperture NA = 0.4, infrared polarizer, quartz beam splitter) while the phase is measured using a custom spatially and spectrally resolved broadband interferometer (see supplementary information).

Plasmons, the collective oscillation of free electrons coupled to photons, shrinks the wavelength of light to make it compatible with biological relevant substances (see refs. [31-35] and references therein). The implementation of plasmonic EPs should enable enhanced sensing of analytes. In Fig. 4a the top gold bars in fabricated DP and EP nanosensors were functionalized with anti-Mouse IgG. The gold bars were first coated with a self-assembled monolayer (linker) by submerging a clean device in an ethanolic solution of 0.1M 8-Mercaptooctanoic acid (MOA) overnight at 4°C. The devices were then activated by standard EDC/NHS (Ethyl-3-(3-dimethylaminopropyl)-carbodiimide/N-hydroxysuccinimide) chemistry. Briefly, the carboxyl ends of the MOA were activated by reaction with EDC (0.4M) and NHS (0.1M) in 4-morpholino ethane sulfonic acid (MES) buffer at pH 6.5 for 35 mins. After drying, the devices were incubated with 100 μL/mL anti-CD63 antibodies for one hour at room temperature. The surfaces were subsequently blocked with 5% bovine serum albumin (BSA) in phosphate-buffered salines (PBS) for 30 mins. After rinsing with PBS, devices were immersed in anti-Mouse IgG with different concentrations overnight at 4°C (figure 4a)

Fabricated samples can only be asymptotically close to the EP singularity without reaching it and can thus only have asymptotically closed eigenmodes (the two modes that coalesce). Resonance splitting of samples covered with the linker is used as a reference and measured ($\Delta\omega_{reference}$). It is presented in Fig. 4b as green horizontal line. The concentration of anti-Mouse IgG in the complex linker/anti-CD63/anti-Mouse IgG (see Fig. 4a) is not proportional to a refractive index. Resonance splitting for EP and DP sensors are thus presented in a histogram. According to Fig. 2g, larger resonance splitting is expected for the EP nanosensor at low enough concentrations. Using the same sample, that is functionalized, cleaned and functionalized again, various concentrations are compared. We observed larger splitting of resonances for the EP nanosensor (red histograms) compared to the DP nanosensor (blue histograms) for concentration smaller than 1 fM ($\Delta\omega_{EP} > \Delta\omega_{DP}$). At 1.25 fM and for higher concentrations, the DP nanosensor is more sensitive than the EP nanosensor ($\Delta\omega_{DP} > \Delta\omega_{EP}$). More interestingly, at 50aM, resonance splitting for the DP sensor is equal to the splitting of the reference ($\Delta\omega_{DP} = \Delta\omega_{reference}$). At the same concentration, resonance splitting of the EP nanosensor is larger than the splitting of the reference ($\Delta\omega_{EP} > \Delta\omega_{reference}$). We further functionalized the EP sample and the DP sample at 30 aM, and Fig. 4b clearly shows that the DP sensors remain at the splitting of the reference while the EP sensors continue to be above the reference.

We have thus experimentally observed exceptional points (EPs) in plasmonics. Plasmonic EPs can be systematically implemented by controlling the interplay between near-field and far-field couplings in hybridized systems governed by Coulomb interactions and interferences respectively. The plasmonic metamaterial EP crystal, made of passive coupled arrays of plasmonic resonators with detuned resonances, exhibit the dispersion of exceptional points around the non-Hermitian singularity and enhanced nanosensing was observed. The ability to drive plasmons to EPs lays the foundation to explore topological physics at small scales and to novel sensors and optoelectronic devices based on topological polaritonic effects.

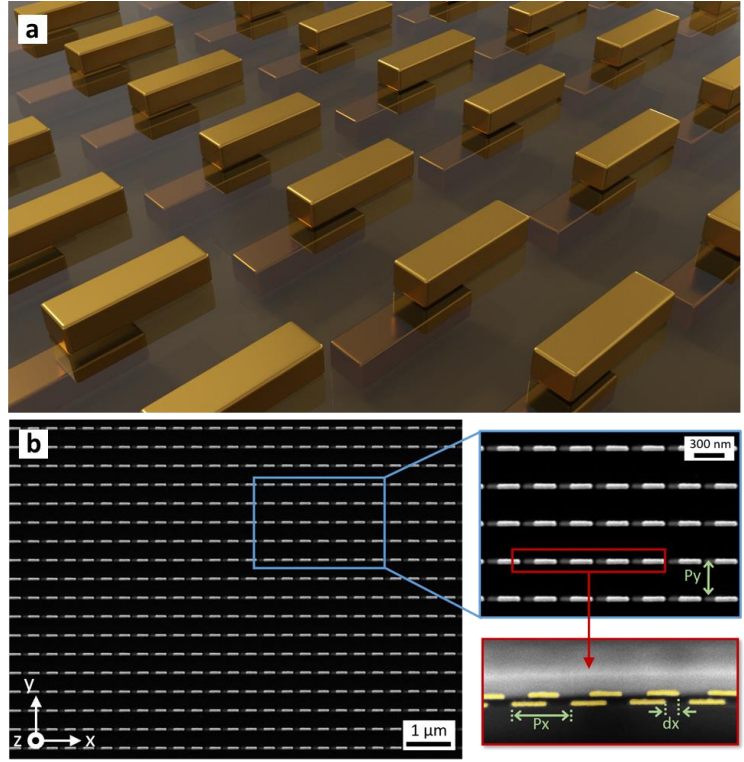

**Fig. 1. Multilayered plasmonic crystal supporting exceptional points.** (**a**) Schematic of a bilayer plasmonic crystal made of two optically dissimilar plasmonic resonators array with detuned resonances. The detuning can be implemented either using structures of distinct size or using identical resonators in distinct optical environments. Here, one of the resonators is on glass and embedded in a 100 nm thick polymer spacer (SU-8) while the second resonator seats on top of the polymer and is exposed to air, making its surface available for analytes binding. The gold metallic bars have a length l = 250 nm, a width w = 50 nm, and a thickness of t = 40 nm. (**b**) Left: Top-view scanning electron micrograph (SEM) of the fabricated multilayer structure, with a lateral shift between bars dx = 100 nm, showing the uniformity of the fabricated structures. Right: Zoom-in top (XY-plane) and side (XZ-plane) views of the plasmonic crystal, clearly showing the top and bottom metallic bars and the quality of the fabrication and alignment processes. The side view image is obtained using a dual-beam focused ion beam (FIB)-SEM that simultaneously enables the local sectioning (with the FIB) and imaging (with SEM) of the samples. Px and Py are in-plane periodicities and dx is the lateral shift between the center of the bars along the direction of their dipolar mode (X-direction). The period along the X-direction is fixed to Px = 400 nm while Py and dx are the two parameters used to reach an exceptional point. The structures are fabricated using two steps electron beam lithography (EBL) and metal lift-off on a glass substrate ($n_{sub}$ = 1.5). The metal array patterns were defined in the bilayer e-beam resists with 170 nm thick methyl methacrylate and 50 nm thick polymethyl methacrylate with high-resolution EBL (EBPG5200, Vistec) followed by a 3 nm chromium and a 37 nm gold deposition using electron

beam evaporation and a lift-off process. The chromium layer is used as an adhesion layer between glass and gold. Then, a 100 nm thick SU-8 2000.1 (MicroChem) is spun over on the first metal array as a planarized dielectric layer ($n_{SU-8}$ = 1.57). The roughness of the surface after planarization is under 3 nm (see supplementary information). Finally, the second layer was fabricated using the same method but including a precise alignment process.

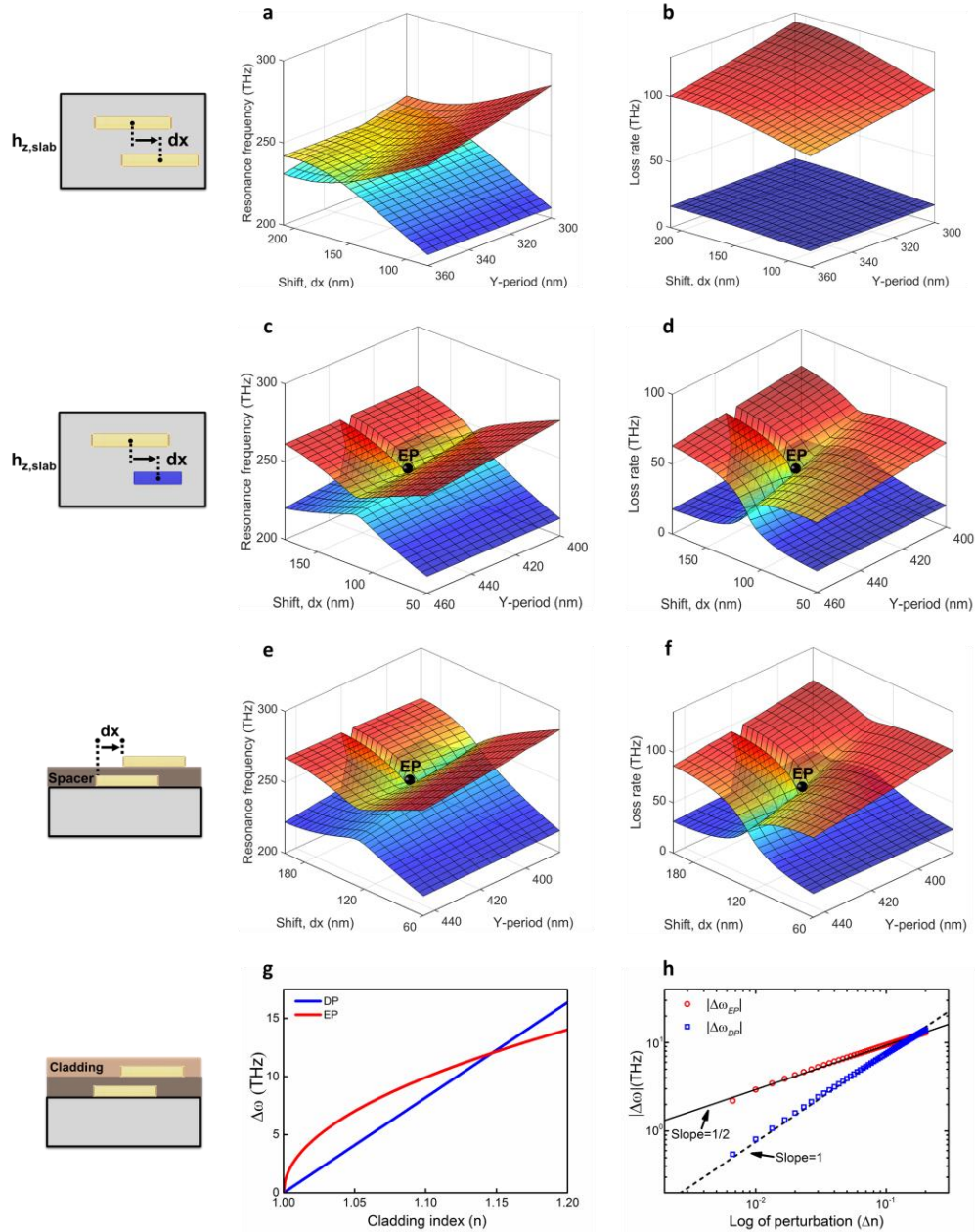

**Fig. 2. Plasmonic exceptional point and symmetry-dependent hybridization scheme of resonances and loss rates.** Real and imaginary parts of the eigenmodes of hybridized plasmonic resonators array of identical size (**a-b**), dissimilar size embedded in a dielectric slab ($n_{slab}$ =1.5 and $h_{z,slab}$ = 240 nm) (**c-d**), and of identical size in an asymmetric optical environment (**e-f**) as a function of the lateral shift between the center of the electric dipoles (dx) and the periodicity perpendicular to the electric dipole moment (Py). The hybridization of optically identical resonators leads to symmetric and antisymmetric modes, and, resonances cross along a diabolic line (**a**) as a function of dx and Py. Because of the opposite symmetry of the hybridized modes, their loss rates (**b**) are very different and are always avoided. The symmetric configuration can thus not lead to exceptional points (EPs). The Hybridization of optically dissimilar resonators, however, leads to two hybrid modes with crossing and avoided crossing of both the resonances and loss rates (**c-d, e-f**) unambiguously demonstrating the existence of a plasmonic exceptional point where resonances and loss rates become simultaneously degenerate. In the last two configurations, the resonators are slightly detuned, but the last configuration is more favorable for sensing as the top metal layer is exposed to air for functionalization and analytes binding. It is worth noting that the loss rate for plasmonic EPs includes losses by radiation and absorption. An EP singularity (black dot) occurs at ~ 241 THz for dx ~ 120 nm and Py ~ 433 nm in (c-d) and at ~ 246 THz for dx ~ 134 nm and Py ~ 415 nm in (e-f). Numerical calculations of scattering parameters are performed using finite element method (see supplementary information). The permittivity of gold is described using a Drude model with a plasma frequency $\omega_p$ = 1.367x10$^{16}$ rad/s and a collision frequency $\omega_c$ = 6.478x10$^{13}$ rad/s. The refractive index of the slab is 1.5. (**g**), Numerical simulations showing resonance splitting in nanostructures supporting a plasmonic exceptional point (EP) and a diabolic point (DP) with a cladding layer described by a refractive index varying from 1 to 1.2. (**h**), Dependence of the logarithm of resonances splitting $|\Delta\omega_{EP}|$ (red circle) and $|\Delta\omega_{DP}|$ (blue square) on the logarithm of the perturbation of (Δn) with slopes of 0.5 and 1.0 respectively. The nanosensor operating at an EP (red line) thus exhibits resonance splitting (Δω) proportional to the square-root of the perturbation **($\boldsymbol{\Delta\omega_{EP}} \sim \sqrt{\boldsymbol{\delta}}$)** whereas the DP nanosensor (blue line) exhibits resonance splitting that depends linearly on the perturbation $\boldsymbol{\Delta\omega_{DP}} \sim \boldsymbol{\delta}$. The square root topology constitutes additional evidence of the successful implementation of a plasmonic EP.

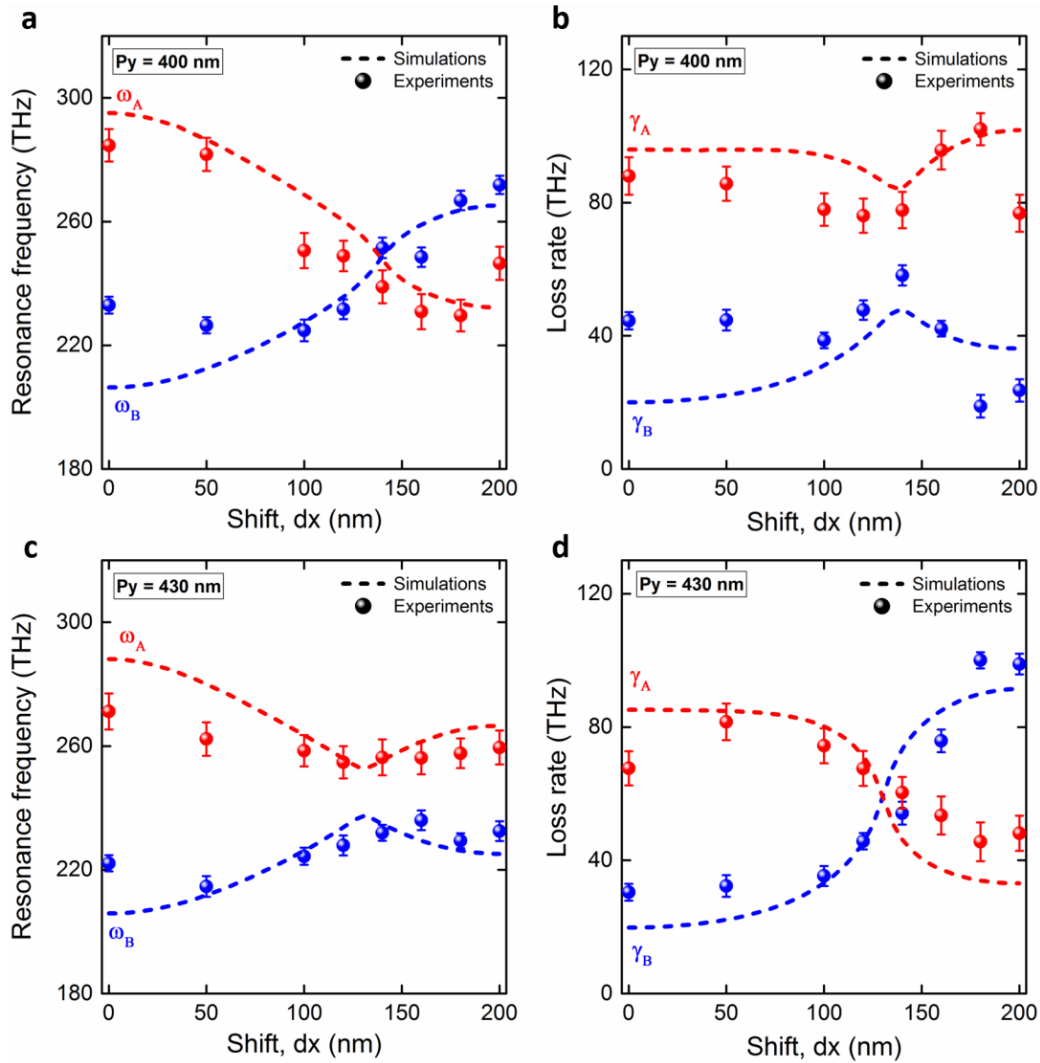

**Fig. 3. Experimental observation of plasmonic exceptional point.** Experimental (circles) and simulated (dashed curves) resonance frequencies (ω) and loss rates (γ), extracted from complex scattering parameters of the structure of Fig. 1, as a function of the lateral shift dx for two different values of the period in the Y direction, Py = 400 nm and Py = 430 nm. (**a, b**) Crossing of the resonance frequencies ($\omega_A$ and $\omega_B$) and avoided crossing of the loss rates ($\gamma_A$ and $\gamma_B$) for Py = 400 nm. (**c, d**) Avoided crossing of the resonance frequencies ($\omega_A$ and $\omega_B$) and crossing of the loss rates ($\gamma_A$ and $\gamma_B$) for Py = 430 nm. In this configuration, an EP singularity thus occurs at ∼243 THz for dx of ∼134 nm and for Py between 400 nm and 430 nm. The experimental results are obtained by measuring both amplitude and phase of the transmitted light. The transmittance is measured using a Fourier-transform infrared spectrometer (Bruker Vertex 70) combined with an infrared microscope (×15 Cassegrain objective, numerical aperture NA = 0.4, infrared polarizer, quartz beam splitter) while the phase is measured using a custom spatially and spectrally resolved broadband interferometer (see supplementary information).

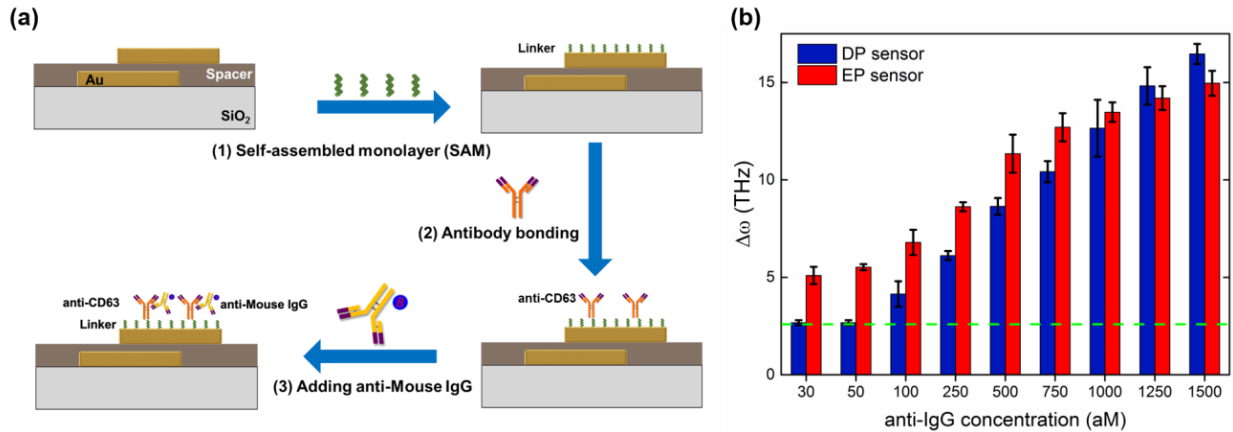

**Fig. 4. Immuno-assay nanosensing with a plasmonic exceptional point.** (**a**) The top bars of the nanosensors are functionalized using the following protocol. In step (1), the sensor was immersed in an ethanolic solution to form a self-assembled monolayer (SAM) denoted linker. In step (2) anti-CD63 was immobilized at room temperature on the sensor. After rinsing with phosphate-buffered salines (PBS), the sensor was immersed [step (3)] in anti-Mouse IgG with different concentrations overnight at 4°C. (**b**) Histograms presenting the measured resonance splitting for different concentrations of anti-Mouse IgG for the Diabolic Point (DP) and Exceptional Point (EP) sensors. The resonance splitting for the sample covered with the linker is measured (green horizontal line) and used as reference ($\Delta\omega_{reference}$). Using the same sample, that is functionalized, cleaned and functionalized again, various concentrations are compared. We observed larger splitting of resonances for the EP nanosensor (red histograms) compared to the DP nanosensor (blue histograms) for concentration smaller than 1 fM ($\Delta\omega_{EP} > \Delta\omega_{DP}$). At 1.25 fM and for higher concentrations, the DP nanosensor is more sensitive than the EP nanosensor ($\Delta\omega_{DP} > \Delta\omega_{EP}$). More interestingly, at 50aM, resonance splitting for the DP sensor is equal to the splitting of the reference ($\Delta\omega_{DP} = \Delta\omega_{reference}$). At the same concentration, resonance splitting of the EP nanosensor is larger than the splitting of the reference ($\Delta\omega_{EP} > \Delta\omega_{reference}$). We further functionalized the EP sample and the DP sample at 30 aM, and Fig. 4b clearly shows that the DP sensors remain at the splitting of the reference while the EP sensors continue to be above the reference.